\documentclass[a4paper,11pt]{article}
\usepackage{pos}
\usepackage{siunitx}
\usepackage{aas_macros}
\usepackage[export]{adjustbox}

\title{Rapporteur Talk: Cosmic Ray Direct}
 \ShortTitle{CRD rapporteur}

\author*[a]{Philipp Mertsch}

\affiliation[a]{Institute for Theoretical Particle Physics and Cosmology (TTK),\\
RWTH Aachen University, 52056 Aachen, Germany}

\emailAdd{pmertsch@physik.rwth-aachen.de}

\abstract{
This is the report on the cosmic ray direct track of the 37$^{\rm{th}}$ International Cosmic Ray Conference (ICRC 2021), broadly covering the contributions relating to charged cosmic rays (CRs) of Galactic origin. 
The contributions highlighted here are of both observational and theoretical nature and aim at interpreting the local fluxes of CRs as well as studying the wider dynamical effects of CRs. 
New data from space-borne experiments on CR electrons and positrons, proton and helium as well as heavier nuclei and their isotopes are reviewed. 
We cover some models of CR acceleration and their feedback on Galactic scales. 
Diffuse emission in $\gamma$-rays as far as it concerns the CR spectra in the immediate vicinity of the solar system and CR anisotropies are covered briefly. 
Some upcoming and proposed experiments are highlighted towards the end. 
}

\FullConference{37$^{\rm{th}}$ International Cosmic Ray Conference (ICRC 2021)\\
		July 12th -- 23rd, 2021\\
		Online -- Berlin, Germany}

\begin{document}
\maketitle

% ----------------------------------------------------------------------------------------
% ----------------------------------------------------------------------------------------
% ----------------------------------------------------------------------------------------
\section{Introduction}

Cosmic rays (CRs) are messengers of the violent universe and interpretations of CR measurements can be used to infer the properties of such environments as far as overall energetics, structure and dynamics are concerned. 
In addition, CRs have a dynamical effect in a variety of environments, shaping not only the medium that they themselves travel in, but also affecting other constituents, e.g.\ magnetic fields, radiation and the thermal gas. 
These two roles of CRs as spectators and of actors allow us to pose different kinds of questions. 
For the former role, the most pressing questions are: What are the sources of CRs? Can we find dark matter in CRs? Is there primordial anti-matter? 
For the latter role, different effects of CRs can be explored: 
CRs produce diffuse emission, contribute to ionisation and heating, provide gravitational support, drive winds and generate turbulence. 
Suffice it to say, what constitutes a valid and interesting model various strongly between these two different directions of research. 

The spectrum of CRs extends over more than 12 orders of magnitude in energy, from tens of MeV to about \SI{10^{20}}{\eV}. 
This spectrum is largely a featureless power law, but there are two important spectral breaks around \SI{\sim 3 \times 10^{15}}{\eV} and \SI{\sim 5 \times 10^{18}}{\eV}. 
Both features have been argued to constitute evidence for the transition from Galactic to extra-galactic sources to take place at either energy. 
At energies than \SI{\sim 10}{MeV/} per nucleon the flux is dominantly of solar origin. 

At International Cosmic Ray Conferences (ICRCs), the scientific contributions on charged CRs have traditionally been distinguished by the kind of detection technique that is used in the respective energy regimes. 
To a certain degree, this distinction falls along the lines of Galactic vs. extra-galactic. 
Up to hundreds of TeV, direct measurements are possible, that is a particle detector, oftentimes consisting of (at least) trackers, calorimeters and scintillation sensors is flown on a balloon or in space. 
This allows not only for determining the energy of an event, but also of the charge and ideally the mass on an event-by-event basis. 
Of course, ideally, direct measurement are desirable at all energies. 
However, given the quick, $\sim E^{-3}$ decrease of (differential) intensities with energy $E$, the number of events for the areas/volumes typically instrumented becomes too low at energies above a PeV and higher. 

This is the report for the cosmic ray direct (CRD) track of the 2021 edition of the ICRC that was to be held in Berlin, Germany, but was eventually exclusively online. 
Naturally, this cannot be an exhaustive review of all contributions to the CRD track, and is unavoidably somewhat subjective. 
Having said that, we tried to cover the most important experimental results and the most controversially discussed topics. 
We will keep references to works other than contributions from this ICRC to a minimum. 
Instead, we refer the interested reader to a few of the recent reviews on the field~\cite{Blasi:2013rva,Kachelriess:2019oqu,Gabici:2019jvz}, in addition to the classical texts, e.g.~\cite{Ginzburg:1990sk}. 
Reports on the other tracks are found elsewhere~\cite{AbuZayyad043,Caputo045,Mitchell:2021mrb,Taoso:2021gvk,Nelles:2021pxq,Strauss:2021oac,Tamborra050,Burton:2021wou}. 

Usually at ICRCs, most contributions are presented in parallel during short oral presentations with limited amount for questions and discussions, in addition to parallel sessions with review and highlight talks. 
Given the exclusive online nature of this ICRC, the scientific content at this year's conference was organised differently. 
Most oral presentations were pre-recorded and made available to all participants ahead of time. 
This allowed significantly more time for discussion in a number of live discussion sessions. 
Overall, it is this rapporteur's opinion that this concept was a great success. 
This is partly due to the minute planning on part of the local organising committee;
but maybe even more so thanks to the tireless efforts of the session conveners who managed to assemble an interesting selection of contributions and trigger stimulating discussions. 
It is our understanding that the recorded contributions, discussion sessions and plenary talks will constitute a semi-permanent record that everybody is invited to peruse\footnote{\url{https://icrc2021-venue.desy.de/}}. 

This report will broadly consist of two parts, in turn addressing these two kind of viewpoints towards CRs mentioned above, viz.\ passive and active, that is CRs as spectators and CRs as actors. 
In Sec.~\ref{sec:spectators} we will review the current observational status, as far as CR direct observations are concerned, ordered by species.  
We will cover new ideas and challenges in the interpretation of these data in parallel. 
A short discussion towards the end of Sec.~\ref{sec:spectators} will highlight some of the remaining discrepancies between observations. 
The second, but much shorter part, Sec.~\ref{sec:actors}, will be almost exclusively devoted to interpretation and modelling of the feedback of CRs in different environments and on a variety of spatial and temporal scales. 
Measurements of CRs through diffuse emission and CR anisotropies, of course, are not exclusively the topic of the CRD track, but we will highlight a few, topical contributions in Sec.~\ref{sec:very_local} and~\ref{sec:anisotropies}, respectively. 
We will briefly cover some of the proposed and upcoming experiments in Sec.~\ref{sec:future} and summarise and conclude in Sec.~\ref{sec:summary}.

% ----------------------------------------------------------------------------------------
% ----------------------------------------------------------------------------------------
% ----------------------------------------------------------------------------------------
\section{Cosmic rays as spectators}
\label{sec:spectators}

% ----------------------------------------------------------------------------------------
% ----------------------------------------------------------------------------------------
\subsection{Electrons and positrons}
\label{sec:electrons_and_positrons}

We start our tour of the latest direct measurements with CR electrons and positrons. 
These have received increased attention over the last couple of years, with positrons in particular suggested as a promising target for searches for particle dark matter (DM). 
We refer the interested reader to the DM rapporteur's talk~\cite{Taoso:2021gvk}.

% ----------------------------------------------------------------------------------------
\subsubsection{Measurements}

The measurement by PAMELA~\cite{PAMELA:2008gwm} of the positron fraction, that is the ratio of positron flux $\phi_{e^+}$ to all-electron flux $(\phi_{e^+} + \phi_{e^-})$ for the first time showed unambiguous evidence of an excess in positrons over pure secondary production by spallation in the interstellar medium (ISM). 
The observational status as of the end of the last ICRC was as follows~\cite{Sparvoli:2020qsg}: 
CR electrons follow a spectrum somewhat softer than $E^{-3}$ at \SI{}{\giga\eV} energies. 
This is shown by the very precise measurements of AMS-02~\cite{AMS:2019iwo}, also confirming the harder than expected positron flux above a few \SI{}{\giga\eV}~\cite{AMS:2019rhg}. 
The calorimetric \textit{Fermi}-LAT experiment also had provided their measurements of the all-electron flux~\cite{Fermi-LAT:2017bpc} and studied electrons and positrons separately by using the Earth's magnetic field. 
At around \SI{\sim 1}{TeV} there is a break to a much softer spectrum, close to $E^{-4}$. 
This feature had already been reported by the H.E.S.S. collaboration~\cite{Aharonian:2008aa,KerszbergICRC2017} and VERITAS~\cite{VERITAS:2018iqb}, that is two \emph{indirect} measurement.
Subsequently, it had been confirmed by DAMPE and CALET, the first \emph{direct} measurements at these energies. 
In addition, DAMPE had found a very narrow feature at \SI{\sim 1}{TeV} which had caused some speculation about an exotic, that is DM origin. 
Note, however, that above \SI{\sim 30}{\giga\eV} there is some disagreement in the spectra between \textit{Fermi}-LAT and DAMPE on the one side and AMS-02 and CALET on the other side. 
\textit{Fermi}-LAT and DAMPE predict a somewhat harder spectrum, that is up to \SI{\sim 40}{\%} higher than the spectrum found by AMS-02 and CALET. 
The origin of this disagreement is, as of yet, unclear. 

\begin{figure}
\centering
\includegraphics[scale=1]{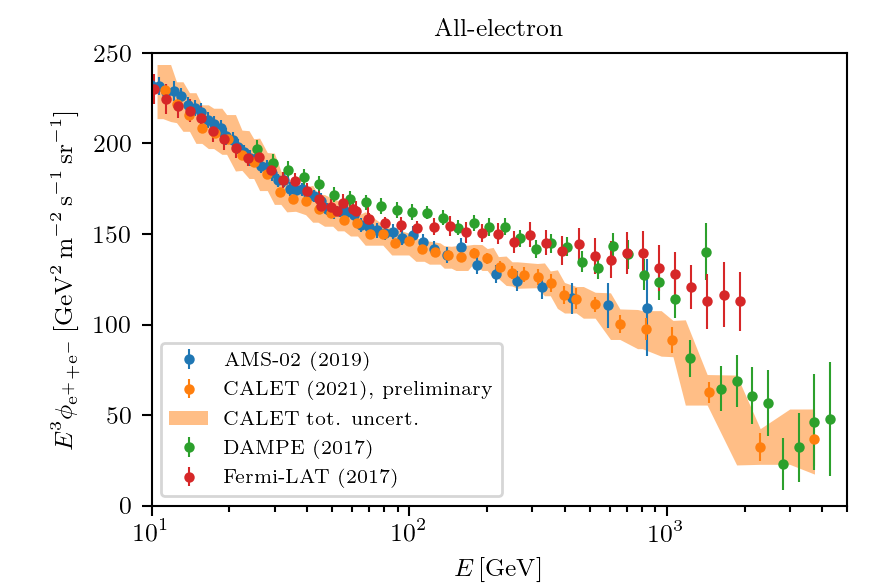}
\caption{Compilation of all-electron fluxes from AMS-02, CALET, DAMPE and \textit{Fermi}-LAT. For the CALET data~\cite{Torii:2021vjo}, the error bars are statistical only, but the shaded yellow region shows the combination of statistical and systematic errors. Other data from the Cosmic Ray Database~\cite{2014A&A...569A..32M}.}
\label{fig:all_electron}
\end{figure}

At this year's ICRC, only the CALET collaboration presented new data~\cite{Torii:2021vjo} on the all-electron flux, see Fig.~\ref{fig:all_electron}. 
Their measurement has significantly improved statistics by a factor of 2.3 over their 2018 analysis. 
The suppression of the flux beyond \SI{\sim 1}{TeV} has now been detected with a significance of $6.5 \, \sigma$. 
As of yet, there is no preference for a broken power law or a power law with an exponential cut-off, but the agreement with DAMPE, the only other direct measurement extending to multi-TeV energies, is good. 

% ----------------------------------------------------------------------------------------
\subsubsection{Interpretation}
\label{sec:interpretation_electrons_positrons}

On the modelling side, there is continuous interest in the electron flux, mostly for its connection to the positron puzzle (see above) and the possibility to associate spectral feature at hundreds of GeV with individual sources. 
The contributions to this modelling discussed at this ICRC can be classified into two groups: those that explain the flux between hundreds of GeV up to a few TeV as dominated by a single source and those that explain it by a population of sources. 
Among the single source scenarios, the objects considered comprise the Vela SNR~\cite{Motz:2021rib} and a potential, unknown, nearby source~\cite{Recchia:2021lcf,Fornieri:2021rol}. 
Among the population scenarios, the objects considered are pulsar wind nebulae~\cite{Linden:2021sky,Orusa:2021ekk,Donato:2021vql}, intrabinary shocks of compact binary millisecond pulsars~\cite{Linares177}, and old supernova remnants (SNRs)~\cite{Mertsch144}. 

A comment is in order on the modelling of CR fluxes from a single source or from an ensemble of sources, either based on a Monte Carlo simulation or some catalogue of observed sources. 
Some models split the ensemble of sources of a particular class, e.g.\ SNRs, into a nearby and a far component. 
Whereas the far sources are modelled as a smooth density, the nearby sources are assumed to be known, e.g. from catalogues. 
However, most likely and for SNRs almost certainly, the nearby sources \emph{cannot} be known from observations to sufficiently large ages. 
It has been shown in a Monte Carlo approach~\cite{Mertsch:2018bqd} that this can lead to an underestimate of the predicted intensities by up to \SI{20}{\%} between a few GeV and few hundred GeV. 
Compared to the observational errors of a few percent, this is a sizeable effect. 
Another fallacy can be committed in the Monte Carlo approach. 
Generally speaking, the spectrum should be rather smooth in energy ranges where a large number of sources contribute and should be fluctuating with energy where only a few sources contribute, for instance due to efficient radiative losses. 
In some models, however, only a small number of sources contribute in energy ranges where the spectrum is very smooth. 
In a Monte Carlo approach this requires some fine-tuning of the spatial and temporal distribution of the sources and their properties. 
In turn, this issue would be reflected in a small fraction of all random draws leading to an acceptable goodness of fit as, e.g.\ quantified by the $\chi^2$ with respect to the data. 
Some authors seem to have adopted a stance along the lines of ``the data is what it is'', but to us it seems obvious that a strongly finely tuned model must be rejected.

Note that most of the interpretation of electron and positron data is done in the framework of conventional diffusion models, that assume a one-zone setup, typical residence times at GeV energies of $\mathcal{O}(10) \, \text{Myr}$ and radiative losses in \SI{}{\mu{}G} magnetic field and radiation field with \SI{}{\eV/\cm^3} energy densities. 
It has been suggested~\cite{Lipari:2021edz}, however, that alternative models which allow for significantly different energy loss time scales and residence times would be able to explain certain spectral similarities that in the standard picture must remain coincidences. 
A particularly interesting coincidence might be the similarity of protons, positrons and anti-protons in spectral shape; 
another one the fact that the ratio of positron and anti-proton fluxes is similar to the ratio of their production cross-sections. 
Alternative scenarios try to explain both positrons and antiprotons as coming from the same source. 
In particular, it might be possible to explain the positron spectrum as purely secondary. 
Alternative models thus entertain the possibility that due to a reduced energy loss rate or increased residence time, CR electrons and positrons do not suffer energy losses below TeV energies. 
However, there remain of course some challenges for these alternative scenarios, in particular with explaining secondary nuclei, in the difference in spectral shapes of positrons and electrons at \SI{\sim 1}{TeV} and with the different source spectra required for electrons and nuclei.

% ----------------------------------------------------------------------------------------
% ----------------------------------------------------------------------------------------
\subsection{Modelling of acceleration}

As it turns out, even the conventional models might require source spectra $\propto \mathcal{R}^{-\gamma}$, with a spectral index different from the canonical $\gamma = 2$ of diffusive shock acceleration. 
Here, $\mathcal{R} \equiv p c / (Z e)$ denotes the rigidity, that is the ratio of momentum $p$ and charge $Z$. 
In the test particle limit, this spectral shape results from the repeated scattering of particles across the shock. 
The isotropisation of the particle distribution after each scattering induces a small, but systematic gain in momentum, the result of which is the formation of a power law spectrum. 
The spectral index is set by the ratio of relative momentum gain and the probability of escaping downstream of the shock. 
If the scattering centres are embedded in the flow, the compression ratio $r$ of the flow, that is the ratio of upstream and downstream speed, solely determines the spectral index as $\gamma = (r+2)/(r-1)$.
For an unmodified, hydrodynamical, strong shock, the compression ratio is $r=4$, thus resulting in the canonical $E^{-2}$ spectrum. 
However, this seems at odds with what can be inferred from the locally measured spectra of CRs with intensities $\phi \propto \mathcal{R}^{-2.8}$. 
At energies where diffusive escape with a diffusion coefficient $\kappa(\mathcal{R})$ dominates, the source spectrum $q(\mathcal{R})$ gets softened to a steady-state spectrum $\propto q(\mathcal{R}) / \kappa(\mathcal{R})$. 
With the rigidity-dependence of the diffusion coefficient $\propto \mathcal{R}^{0.3 \mathellipsis 0.5}$ as inferred from measurements of nuclear secondary-to-primary ratios, e.g.\ boron-to-carbon, this can only be explained by a source spectrum $\propto \mathcal{R}^{-2.5 \mathellipsis -2.3}$, thus much softer than the canonical $\mathcal{R}^{-2}$. 
Note that this discrepancy usually gets worse if the feedback of CR pressure on the velocity profile is taken into account, due to the predicted spectra being even harder for such scenarios.
See Ref.~\cite{Malkov:2001kya} for a review. 

As far as explanations for the softer source spectra go, a number of possibilities have been considered on the theoretical side of the CRD track. 
% ----------------------------------------
% M. Pohl #987
% ----------------------------------------
An answer in the negative has been provided to the suggestion that the softening is due to the energy loss that ions suffer when generating turbulence through the so-called Bell instability. 
In particular, it was shown~\cite{Pohl:2021lct} that the steady-state picture assumed in such arguments does not apply and that once the finite size of the shock precursor is taken into consideration, the change in spectral index can only be 0.1 under most favourable conditions. 
% ----------------------------------------
% S. Das #988
% ----------------------------------------
Considering realistic velocity profile as encountered, for instance, by a blast wave expanding into the shocked wind of a massive progenitor star, might be one way to go beyond the standard scenario sketched out above. 
In this case, the compression ratio at the shock will be time-dependent, but at most times, it will be smaller than $r = 4$, sometimes even as small at $r = 1.5$. 
The integrated spectrum of shock accelerated particle will therefore quite naturally be softer~\cite{Das:2021fux}. 
% ----------------------------------------
% D. Caprioli #482
% ----------------------------------------
Finally, it can also be important to take into account that the scattering centres that the CRs interact with have a finite velocity in the fluid frame. 
In particular, on the downstream side this will increase the effective velocity such that the effective compression ratio experienced by the CRs will be smaller than the hydrodynamical $r = 4$. 
It was argued~\cite{Caprioli:2021nwv} that the additional speed is in fact the Alfv\'en speed. 
Softer spectra have indeed be observed in particle-in-cell (PIC) simulations, but whether the additional speed can be identified with the Alfv\'en speed of the turbulence generated by the Bell instability is a different matter; generally this turbulence has been found to be non-Alfv\'enic.

\begin{figure}
\includegraphics[width=0.48\textwidth,valign=c]{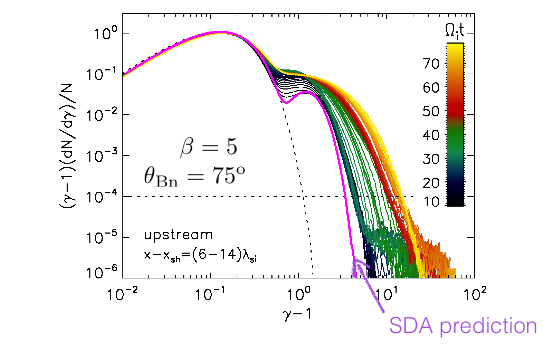}
\includegraphics[width=0.48\textwidth,valign=c]{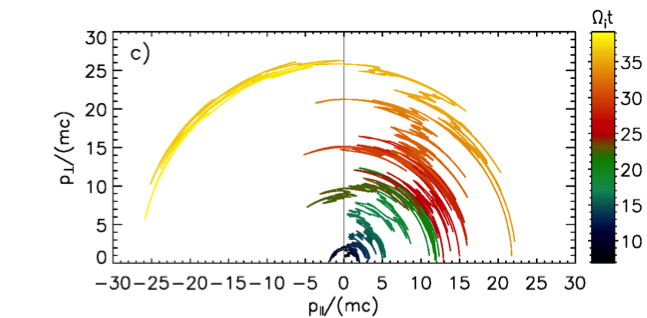}
\caption{Results from PIC simulations of stochastic SDA at a quasi-perpendicular shock. \textbf{Left panel:} The spectra of accelerated particles extended significantly beyond the SDA prediction. \textbf{Right panel:} The distribution in momentum space shows phases of acceleration (movement in radial direction) and pitch-angle scattering (movement in azimuthal direction). From Ref.~\cite{Niemiec:2021okp}.}\label{fig:Niemiec_et_al}
\end{figure}

An important precondition for a particle to be shock accelerated is for the particle gyroradius to exceed the width of the (thermal) sub-shock. 
This condition is generically not satisfied by ions from the thermal distribution, and certainly not for electrons. 
There are however observations of synchrotron and X-ray emission from a variety of systems, that bear witness to the efficient acceleration of CR electrons, even at quasi-perpendicular shocks, an example being cluster merger shocks. 
More broadly, this so-called injection problem has been plaguing the modelling of shock acceleration for a long time. 
More recently, however, it has been suggested that so-called shock drift acceleration (SDA) could provide the necessary mechanism to energise particles from the thermal tail to the point where diffusive shock acceleration can operate. 
% ----------------------------------------
% J. Niemiec #477
% ----------------------------------------
At this ICRC, this was the topic of one discussion session, but we will highlight one contribution in particular. 
In SDA, particles are being accelerated in the motional electric field of the shock. 
However, after a short time, they will generically be reflected away from the shock and will therefore only experience a moderate amount of acceleration. 
However, the flux of reflected particles can trigger an instability, thus producing turbulence that can scatter them back towards the shock, therefore allowing for subsequent cycles of SDA. 
Taken together, this process has been dubbed ``stochastic shock drift acceleration''. 
In PIC simulations, it was shown~\cite{Niemiec:2021okp} that within a few tens of gyrotimes the spectrum can extend significantly in energy such that the threshold for shock acceleration is reached (see the left panel of Fig.~\ref{fig:Niemiec_et_al}), even for highly oblique shocks. 
Most interestingly, the trajectory of particles in momentum space clearly show the alternating between periods of acceleration and periods of pitch-angle scattering during which the absolute value of the momentum remains constant (see the right panel of Fig.~\ref{fig:Niemiec_et_al}). 

PIC simulations are a great tool for studying the development and saturation of plasma instabilities which have invariably been seen to be at work in many environments. 
But due to their high computational expense, PIC simulations cannot currently be applied to the full astrophysical problem. 
In studies of acceleration at non-relativistic shocks, for example, the longest running codes have just covered $\mathcal{O}(10^3)$ gyroperiods. 
It has been argued that after this amount of time, the typical textbook picture of shock acceleration, that is non-thermal particles crossing the shock essentially unimpeded, has not been observed yet and would probably require much longer. 
Extending the times covered to the dynamical time scales of, e.g.\ SNRs would require simulations even longer by orders of magnitude.

\begin{figure}
\includegraphics[width=\textwidth]{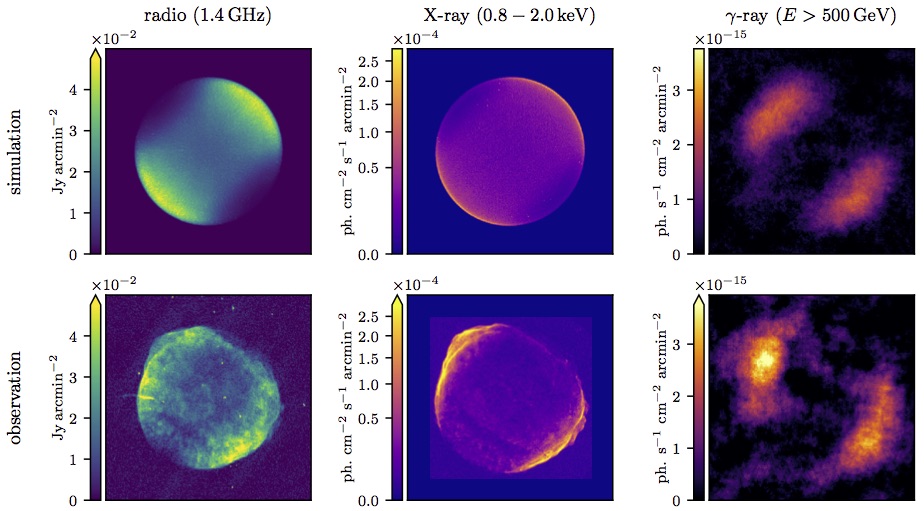}
\caption{Comparison of synchrotron (left column), X-ray (middle column) and $\gamma$-ray (right column) flux maps of SN1006, from a ``PIC-informed'' MHD simulation (top row) and from observations (bottom row). From Ref.~\cite{Pfrommer425}.}
\label{fig:Pfrommer_et_al}
\end{figure}

What is thus needed are ways to ``bridge the gap'' between the first principle PIC simulations and other, effective computational approaches, e.g. for the hydrodynamic, long-time evolution of the system.
At this ICRC, we have been able to identify three kinds of approaches and we will here highlight one exemplary contribution each. 
% ----------------------------------------
% C. Pfrommer #425
% ----------------------------------------
Taking lessons from PIC simulations and implementing those in a heuristic way in, e.g. MHD simulations is what can be called ``PIC-informed'' MHD simulations. 
One example is the study of the morphological and spectral properties of non-thermal emission from SNRs. 
From PIC simulations, it is know that the acceleration efficiency at a non-relativistic shock is a function of the shock obliquity and Mach number. 
As both these macroscopic quantities can be determined from MHD simulations of SNRs, these dependencies can be implemented into MHD simulations in oder to accelerate CRs in the right numbers~\cite{Pfrommer425}. 
In Fig.~\ref{fig:Pfrommer_et_al}, observations are compared to predictions for preferential acceleration of electrons at quasi-perpendicular shocks, as seen in PIC simulations. 
Interestingly, the radio and X-ray morphologies do \emph{not} agree and better agreement would require preferential \emph{quasi-parallel} acceleration.  
The value of such approaches rises and falls, naturally, with the reliability of the recipes adopted. 

% ----------------------------------------
% A. J. van Marle #447
% ----------------------------------------
Another possible approach is to combine MHD simulations for the thermal particles with PIC simulations for the non-thermal ones. 
This is similar in spirit to so-called hybrid PIC simulations; only that here, the non-thermal ions are treated in the magneto-fluid part. 
This technique is known as MHD-PIC.
Thanks to its capacity of running in larger simulation boxes and for longer amounts of time, MHD-PIC can resolve long-wavelength instabilities that pure PIC cannot.
Naturally, this also requires some recipes for under which conditions to move particles from the MHD to the PIC simulation. 
As for the physics results, it was confirmed with this technique~\cite{vanMarle:2021rmb} that even the long-wavelength modes are not capable of accelerating particles at moderate Mach number and high obliquities. 

% ----------------------------------------
% A. Bohdan #443
% ----------------------------------------
Finally, it is worthwhile remembering the heliosphere as a laboratory for studying wave-particle scattering or kinetic instabilities. 
Saturn possesses a high Mach number bow shock that has been observed \textit{in-situ} by the Cassini spacecraft. 
PIC simulations can thus be compared to data~\cite{Bohdan:2021qin}. 
In PIC simulations, the magnetic field amplification has been seen to be due to the Weibel instability. 
The saturation levels reached compare favourably with Cassini data.

% ----------------------------------------------------------------------------------------
% ----------------------------------------------------------------------------------------
\subsection{Heavy nuclei}

Most of the new direct measurements presented at this ICRC came from the AMS-02, CALET, DAMPE or ISS-CREAM collaborations. 
CR proton and helium fluxes are, of course, the most prominent species. 
Due to their high relative abundances, they are measured with the highest statistics and their spectra extend to the highest energies.
We will discuss the new proton and helium measurements in Sec.~\ref{sec:proton_helium}. Some attention has, however, also been devoted to ``heavier nuclei'', that is species with mass number $Z > 2$. 
The AMS-02 experiment, in particular, has made great strides towards providing precision measurements of all nuclei towards iron. 
Before reviewing those new data we will motivate such studies. 

It could be argued that measurements of spectra of species heavier than helium cannot contribute much valuable new information. 
This preconception is based on the belief of CR universality: 
As a collisionless plasma, CRs only scatter off magnetic fields and so their spectra can only depend on rigidity $\mathcal{R}$.
Universality must be broken in energy ranges where other effects are important, e.g.\ nuclear interactions. 
This points to the first application of abundance measurements, that is testing this universality. 
Second, the abundance of the different species contains important information about the environments that the sources of CRs are accelerating particles from.
The second line of reasoning can be traced back ultimately to the seminal work of Meyer, Drury and Ellison~\cite{Meyer:1997vz,Ellison:1997an} who suggested that the source abundances of CR nuclei (that is before contribution from spallation in the interstellar medium) could be understood following two principles: 
First, a charge-to-mass dependence, probably determined by the microphysics of shock accelerations. 
Second, a preference of refractory elements (which are enclosed in dust grains) with respect to volatiles (which a predominantly in the gas phase). 
Here, we will highlight two contributions that have applied and updated this kind of reasoning to modern data. 

\begin{figure}
\centering
\includegraphics[width=0.8\textwidth]{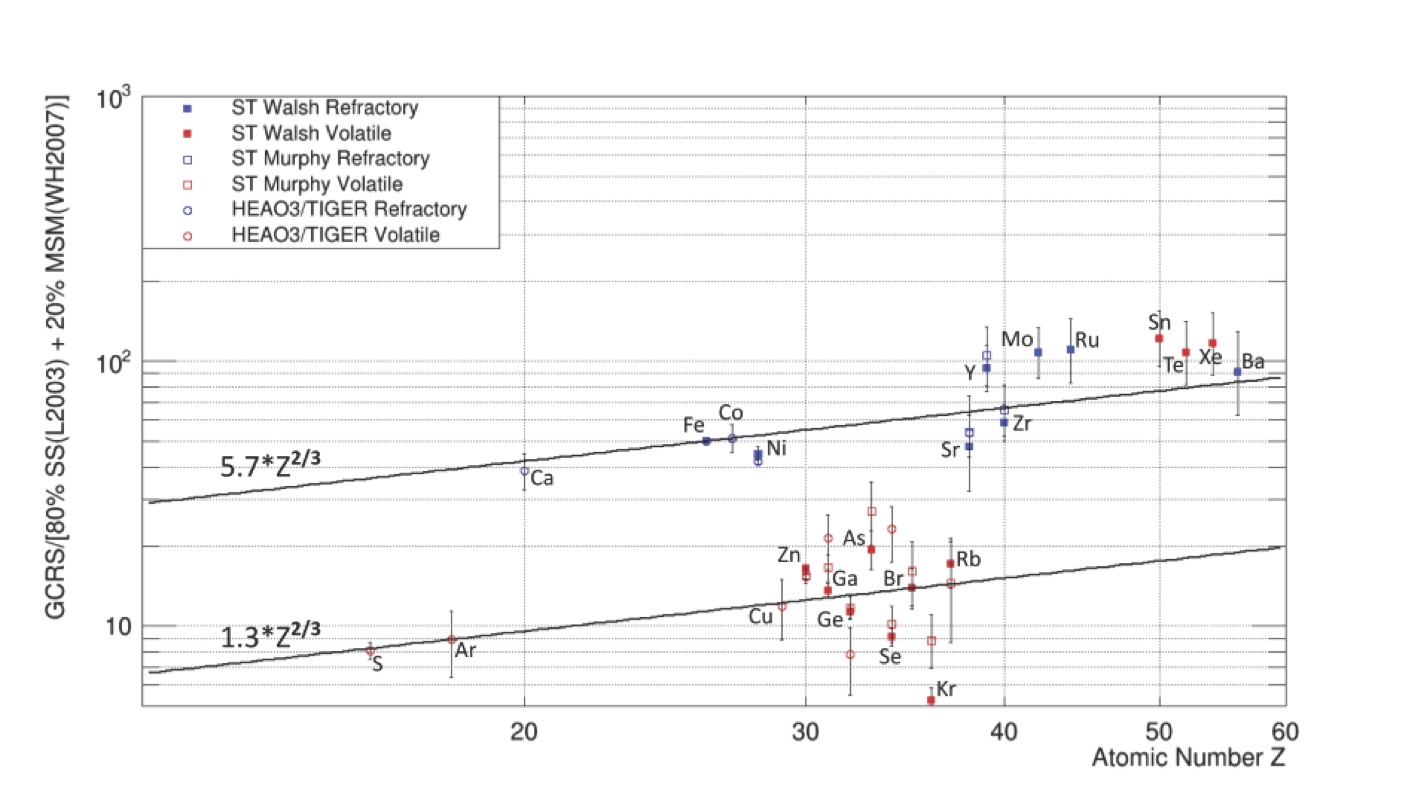}
\caption{Galactic cosmic ray source (GCRS) abundances as a function of atomic number $Z$. The abundances have been normalised to 80\,\% solar system material and 20\,\% ``massive star'' material. Note, how the split between refractories and volatiles, visible for $Z \lesssim 40$, disappears for $Z > 40$. From Ref.~\cite{Walsh:2021fwp}}
\label{fig1}
\end{figure}

% ----------------------------------------
% N. Walsh (Super-TIGER) #118
% ----------------------------------------
At this ICRC, the SuperTIGER collaboration presented new results on their measurements of the abundance of heavy elements up to $A = 56$~\cite{Walsh:2021fwp}. 
Previously, the TIGER experiment had confirmed the split between refractory and volatiles as suggested by Ellison, Drury and Meyer. 
They had shown that the predicted scaling would however only be observed when not assuming solar system abundances, but rather a mix of 80\,\% solar system material and 20\,\% ``massive star'' material. 
This behaviour had been observed up to $Z \simeq 29$; the updated analysis shows, however, that this behaviour does not extend beyond $Z = 40$. 
Instead, beyond $Z = 40$, there is no suppression of volatiles with respect to refractories, that is volatiles with $Z \gtrsim 40$ follow the same $Z$-dependence as refractories of all $Z$. 
This is particularly evident, if only even charge number elements are considered beyond $Z = 40$, see Fig.~\ref{fig1}. 
(The odd charge number ones are more difficult to measure due to their lower \emph{absolute} abundances.) 
While highly speculative, this fact might point to the importance of the $r$-process in the origin of $Z \gtrsim 40$ elements, a hint that might offer interesting connections to the evidence for the $r$-process in neutron star binary mergers. 

% ----------------------------------------
% V. Tatischeff #153
% ----------------------------------------
On the theoretical side, an interesting update of the idea of Meyer, Drury and Ellison was presented~\cite{Tatischeff:2021iwh}, making use of our better understanding of the environments of various candidates of CR sources as far as composition is concerned. 
This builds on the insight that source abundances depend primarily on (1) the composition of the ``source reservoir'', (2) the ISM phase from which CRs are accelerated and (3) the dust content. 
It is found that volatiles must be mostly from superbubbles, with SNRs in the warm ISM contributing less than \SI{30}{\%} overall. 
Refractories can also originate from superbubbles as long as a continuous replenishment of dust is guaranteed. 
It is noteworthy that some of their conclusions sensitively depend on the observation that the inferred source spectra of proton, helium and heavier elements differ significantly from each other. 

% ----------------------------------------
% H. Gast #121
% ----------------------------------------

Shortly after ICRC 2017, AMS-02 had presented their measurements of both primaries helium, carbon and oxygen as well as secondary lithium, beryllium and boron and these results were reviewed at this ICRC~\cite{Gast121}. 
The rigidity spectra of these three primary species agree in shape above \SI{50}{GV} with a prominent break around \SI{200}{GV} from a $\mathcal{R}^{-2.7}$ power law below to a $\mathcal{R}^{-2.6}$ behaviour above. 
The shapes of the three secondaries also agree well with each other and exhibit a break also around \SI{200}{GV}. 
However, this break is about twice as strong, from $\mathcal{R}^{-3.1}$ to $\mathcal{R}^{-2.9}$. 
This fact has been widely interpreted as being due to a propagation effect. 
For instance, if the diffusion coefficient in the interstellar medium hardened by $0.1$ in spectral index around \SI{200}{GV} and if the source spectra were power laws, $q(\mathcal{R}) \propto \mathcal{R}^{-\Gamma}$, the propagated \emph{primary} spectra, $\psi_1 \propto q(\mathcal{R}) / \kappa(\mathcal{R})$ would have a break of $0.1$, too. 
The propagated \emph{secondary} spectra $\psi_2 \propto \psi_1(\mathcal{R}) / \kappa(\mathcal{R}) \propto q(\mathcal{R}) / \kappa(\mathcal{R})^2$ instead were affected twice and had a break of $0.2$ in spectral index. 
Nitrogen exhibits a spectrum reminiscent of secondaries at lower rigidities and of primaries at higher rigidities, thus it must have some source abundances. 

\begin{figure}
\centering
\includegraphics[width=0.7\textwidth]{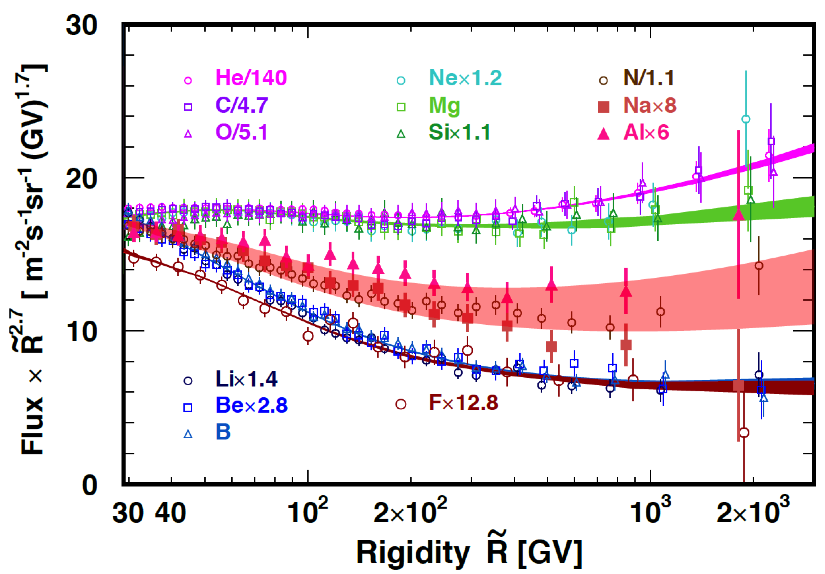}
\caption{Compilation of measurements of heavy nuclei fluxes by AMS-02. From Ref.~\cite{Oliva107}.}
\label{fig:AMS-02}
\end{figure}

% ----------------------------------------
% A. Oliva #107
% ----------------------------------------
On this background, new rigidity spectra of neon, magnesium, silicon, fluor, sodium and aluminium were presented at this year's ICRC~\cite{Oliva107}. 
These measurements, together with the earlier ones are summarised in Fig.~\ref{fig:AMS-02}. 
Neon, magnesium and silicon are well in agreement in their spectral shapes above \SI{100}{GV} and show rather hard spectra $\propto \mathcal{R}^{-\delta_{1,2}}$, both below ($\delta_1 = 2.75$) and above ($\delta_2 = 2.65$) the \SI{\sim 200}{GV} break. 
The spectral shapes are slightly different from the helium, carbon and oxygen ones though, which was said could be due to ``two different classes'' of sources. 
Fluor shows a spectrum typical for a secondary and both sodium and aluminium are again a mixture of secondary and primary contributions. 

% ----------------------------------------
% M. Korsmeier #176
% ----------------------------------------
% M. Vecchi #174
% ----------------------------------------
On the modelling side, this abundance of spectra for different species should make for great interest. 
Apart from the information on source candidates and source environments, each additional species has the potential of reducing uncertainties in fitted transport parameters and provide further checks of the model. 
However, this progress is hampered by our limited knowledge of spallation cross-sections as highlighted by two contributions at this year's ICRC. 
In fact, with the parametrisations as implemented in popular propagation codes like \texttt{GALPROP}, there are residuals in the model fits. 
In light of a lack of a first principles model able to predict these cross-sections and the scarcity of modern cross-section data, phenomenologists have resorted to simple parametrisations of corrections to the cross-sections. 
The parameters of such corrections can be considered nuisance parameters and can be included in the fit of the transport parameters~\cite{Korsmeier:2021nuk}. 
While the goodness of fit is significantly improved, there are degeneracies between nuisance parameters and model parameters. 
A particularly spectacular illustration has been shown for the discrepancies with existing cross-section parametrisations~\cite{Vecchi:2021qca}. 
It was shown that with current cross-sections a diffusion model that has been fit to Li, Be, B, He, C and O data overproduces the fluor-to-silicone (F/Si) ratio as measured by AMS-02 by \SI{20}{\%}. 
If instead the cross-section is modified in normalisation by a similar amount, the F/Si data and model predictions largely agree. 

% ----------------------------------------
% N. Amin #102
% ----------------------------------------
What is needed are thus more accurate cross-section parametrisations which need to be informed by more cross-section measurements. 
The only experimental effort in this direction that was presented at this ICRC was by the NA61/SHINE collaboration~\cite{Amin:2021oow}. 
In a pilot run in 2018 at $13.5 \, A \, \text{GeV}$ beam energy, $A$ being the nuclear mass number, the spallation of proton on carbon was studied by combining measurements with a polyethylene ($\text{C}_2\text{H}_4$) target and a graphite ($\text{C}$) target. 
The results agree largely with earlier data, which goes towards showing the reliability of such studies. 
While the error bars of this pilot run are still of similar size as those of earlier measurements, this will be improved on in future runs. 
Currently, data taking for light secondaries (B, Li and Be) production on light primaries (C, N and O) is scheduled for 2022. 

% ----------------------------------------
% L. Derome #992
% ----------------------------------------
An important limitation in our knowledge of the transport of galactic CRs is related to our ignorance of the confinement time of CRs in the extended transport volume. 
In a simple diffusion model in which the CR sources and the gas are confined to the disk, there is a degeneracy between the normalisation of the diffusion coefficient $\kappa$ and the height of the CR halo, $z_{\text{max}}$. 
In fact, measurements of ratios of \emph{stable} nuclei, e.g.\ the famous boron-to-carbon ratio constrain the combination $(\kappa / z_{\text{max}})$. 
The flux of \emph{unstable} nuclei instead depends differently on the diffusion coefficient and therefore nuclear ratios involving unstable nuclei provide an independent measure of $\kappa$ that allows to break the degeneracy between $\kappa$ and $z_{\text{max}}$. 
Such unstable nuclei have been dubbed ``CR clocks'' and while their importance was recognised early on, the quality of the measurements had left something to be desired in terms of statistics and energy reach. 
A significant improvement over older measurements has been presented at this year's ICRC by the AMS-02 collaboration~\cite{Derome119}. 
The main challenge for this measurement is the limited mass resolution of \SI{\sim 1}{amu} even when combining different subsystems. 
Therefore, the relative abundances must be determined by fitting templates to the observational mass distributions in the individual energy bins. 
A careful analysis has enabled the measurements of $\mathstrut^6\text{Li}$ and $\mathstrut^7\text{Li}$ and fluxes between \SI{0.3}{} and \SI{11}{\giga\eV\per n} and of $\mathstrut^7\text{Be}$, $\mathstrut^9\text{Be}$ and $\mathstrut^{11}\text{Be}$ between $0.5$ and $12 \, \text{GeV/n}$. 
The measurements of the isotopic ratios, e.g.\ $\mathstrut^{10}\text{Be}/\mathstrut^9\text{Be}$, have significantly extended the energy reach and precision over previous measurements. 
Most interestingly, $\mathstrut^{10}\text{Be}/\mathstrut^9\text{Be}$ remains almost constant and close to \SI{0.2}{} between \SI{1}{} and \SI{10}{\giga\eV\per n} and no sharp upturn is observed, which should point to a rather large height of the CR halo. 
Apart from the scientific significance, this analysis stands out since not only errors but also correlation matrices are provided. 
This should prove most useful for a sound interpretation of the data.

% ----------------------------------------------------------------------------------------
% ----------------------------------------------------------------------------------------
\subsection{Discrepancies}

Of the other current, space-borne experiments, CALET presented recent measurements of carbon, oxygen and iron fluxes~\cite{Maestro:2021gwg,Stolzi:2021ern}. 
The carbon and oxygen flux measurements extend in kinetic energy per nucleon between \SI{\sim 10}{GeV/n} and \SI{\sim 2}{TeV/n}. 
While the shapes are close to $E^{-2.7}$ below a few hundred GeV/n, the spectra get markedly harder above. 
While the spectral shapes are in agreement with those measured by AMS-02, the normalisations found by CALET are lower by about $27 \, \%$. 
The CALET C and O measurements are however in agreement with PAMELA data. 
The origin of this discrepancy is as of yet unknown. 
Note that AMS-02 and CALET (as well as PAMELA) agree on the flux ratio C/O. 
Finally, we note that the normalisation of the boron flux measured by CALET is similarly lower than found by AMS-02 while the boron-to-carbon ratios agree~\cite{Akaike:2021xsl}. 
A similar comment applies to the iron measurements, recently presented by AMS-02~\cite{Chen:2021xyw} and CALET~\cite{Stolzi:2021ern}. 

The discrepancies between CALET and AMS-02 are all the more surprising given the good agreement of their measurements of the all-electron flux (see Sec.~\ref{sec:electrons_and_positrons} above). 
We are thus left in the rather peculiar situation in which the all-electron flux measurements agree between CALET and AMS-02, both in shape and normalisation while the primary fluxes C, O and Fe disagree. 
It was pointed out in one of the discussion sessions that this disagreement in the nuclear normalisations might not be that surprising, given the respective calorimeters' depths in interaction lengths. 
It thus seems clear that further work is needed by either collaboration to resolve this issue. 
As both collaborations are adamant that their respective measurement is correct, it is difficult to see how this conflict could possibly be resolved. 
To this rapporteur, the situation bears some resemblance with the disagreement between the AUGER and TA collaborations on the all particle ultra-high energy CR spectrum. 
In this case, some agreement could be found by a working group composed of members of both collaborations~\cite{TelescopeArray:2021zox}. 
Maybe a similar effort between the AMS-02 and CALET collaborations could also help resolving the conflicting results on the C, O and Fe flux measurements.

% ----------------------------------------------------------------------------------------
% ----------------------------------------------------------------------------------------
\subsection{Proton and helium}
\label{sec:proton_helium}

\begin{figure}[tbh]
\centering
\includegraphics[scale=1]{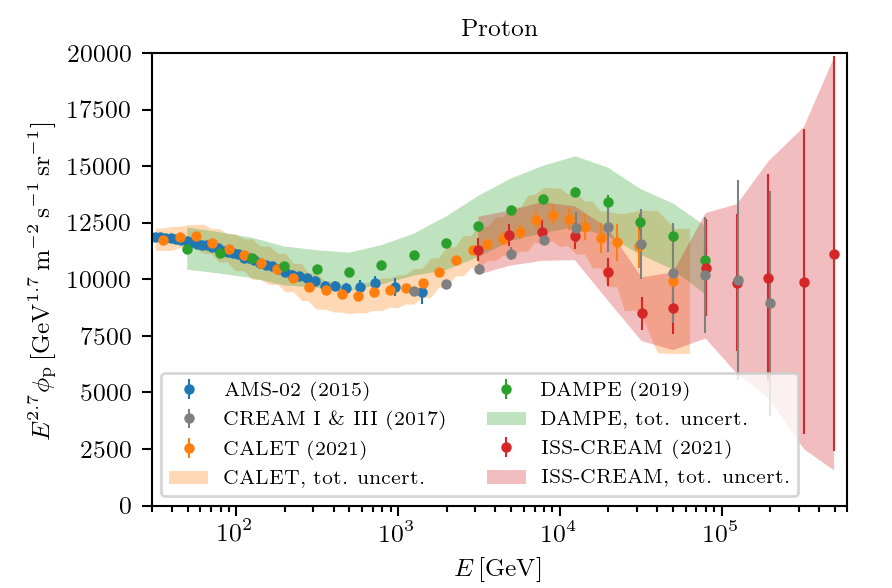} \\
\includegraphics[scale=1]{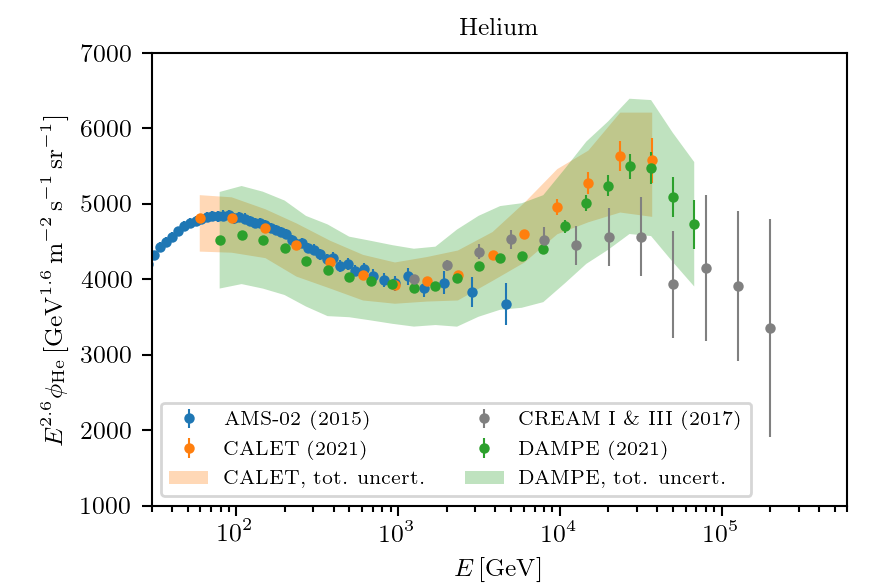} \\
\caption{Proton (top panel) and helium (bottom panel) measurements by AMS-02, CALET, DAMPE and ISS-CREAM as well as ancillary data. Data from Refs.~\cite{Kobayashi:2021zml,Brogi:2021csp,Choi:2021nwe} and the Cosmic Ray Database~\cite{2014A&A...569A..32M}.}
\label{fig:proton_helium}
\end{figure}

Of the four space experiments mentioned above, both CALET and DAMPE have presented proton and helium fluxes at this ICRC. 
Both, the proton (helium) fluxes were seen to have power law spectra $\propto E^{-2.8}$ ($\propto E^{-2.7}$) below and $\propto E^{-2.7}$ ($\propto E^{-2.6}$, possibly harder) above a break at a few hundred GeV (\SI{\sim 1}{TeV}). 
In addition, DAMPE had previously reported a softening at \SI{10}{TeV}. 
% ----------------------------------------
% K. Kobayahsi #98
% ----------------------------------------
At this ICRC, CALET has updated their measurements of the proton flux, extending the energy reach from \SI{10}{TeV} to \SI{60}{TeV}~\cite{Kobayashi:2021zml}, see Fig.~\ref{fig:proton_helium}, top panel. 
The hardening is now located at \SI{\sim 550}{GeV} and the softening in protons observed by DAMPE and indicated by earlier balloon data is confirmed. 
The analysis by CALET puts this feature at \SI{11}{TeV}.
% ----------------------------------------
% X. Li #13
% ----------------------------------------
The DAMPE collaboration had presented their first measurement of the proton flux between \SI{50}{GeV} and \SI{\sim 80}{TeV} shortly after the last ICRC, which was reviewed at this year's conference~\cite{Li13}. 
Their analysis is largely in agreement with CALET, but also with PAMELA, AMS-02, ATIC-2, CREAM and NUCLEON in the respective energy ranges. 
The energy of the hardening and softening breaks in proton were found to be at \SI{\sim 500}{GeV} and \SI{14}{TeV}, respectively, thus also roughly agreeing with the results by CALET. 

% ----------------------------------------
% P. Brogi #101
% ----------------------------------------
CALET also reported its first measurement of the He flux between a few hundred $\text{GeV}$ and \SI{\sim 50}{TeV}~\cite{Brogi:2021csp}, see Fig.~\ref{fig:proton_helium}, bottom panel. 
The energy of the hardening is \SI{1.3}{TeV} which corresponds to a rigidity of \SI{\sim 650}{GV}, somewhat higher than what for instance AMS-02 finds for the rigidity of the spectral breaks in nuclei. 
% ----------------------------------------
% M. Di Santo #114
% ----------------------------------------
The corresponding analysis by DAMPE~\cite{DiSanto:2021wlr} is again in agreement with CALET (as well as previous data: PAMELA, AMS-02, ATIC-2, CREAM and NUCLEON). 
The hardening is observed between 200 and \SI{300}{GeV/n} in kinetic energy per nucleon, thus in agreement with the value inferred by CALET. 
Interestingly, the DAMPE analysis reaches up to \SI{\sim 20}{TeV/n} and has established a softening of the helium spectrum around \SI{34}{TeV}. 
Comparing this to the energy of the softening break seen in protons, that is \SI{\sim 10}{TeV}, both the proton and helium break could be at the same rigidity or at the same energy per nucleon. 

% ----------------------------------------
% G. Choi #94
% ----------------------------------------
Finally, the ISS-CREAM experiment also presented their analysis of the proton flux~\cite{Choi:2021nwe}, see Fig.~\ref{fig:proton_helium}, bottom panel. 
Their measurement spans from \SI{2.5}{TeV} to \SI{655}{TeV}, albeit with very limited statistics above \SI{\sim 50}{TeV}. 
Below, there is a softening at \SI{\sim 12}{TeV}, in agreement with what has been found by DAMPE and CALET. 
Above, the error bars are too large to allow for any reliable interpretation although the data are certainly compatible with another hardening. 

\begin{figure}[tbh]
\centering
\hspace{1.2em} \includegraphics[width=0.61\textwidth,trim={0 0 0 4cm},clip=True]{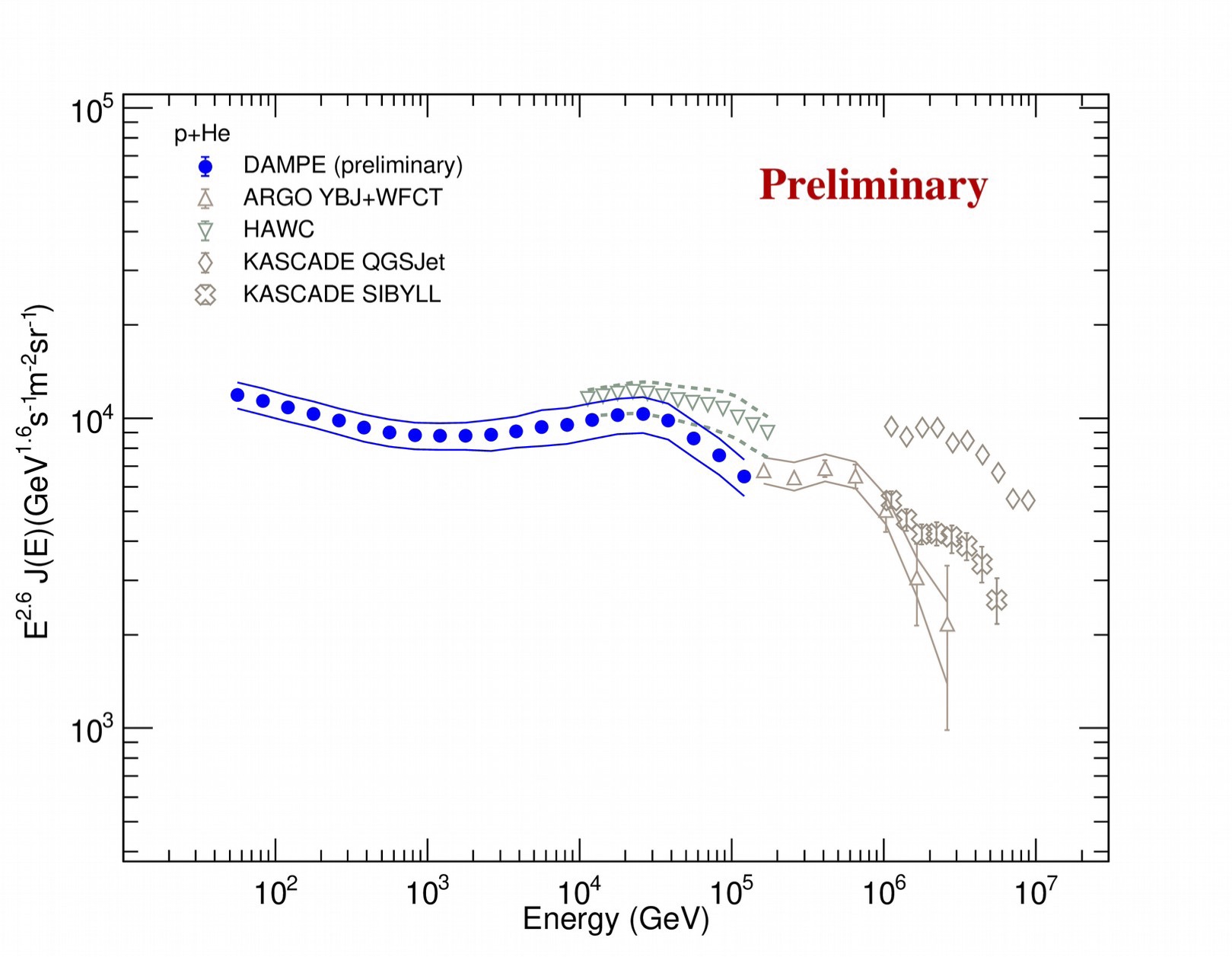}
\caption{Combined proton plus helium flux measured by DAMPE and compared to other measurements. From Ref.~\cite{Alemanno:2021ldn}.}
\label{fig:proton+helium}
\end{figure}

% ----------------------------------------
% F. Alemanno #117
% ----------------------------------------
An interesting way to extend the energy reach beyond what is possible for proton and helium separately was presented also by the DAMPE collaboration~\cite{Alemanno:2021ldn}, see Fig.~\ref{fig:proton+helium}. 
This method enables a direct measurement of the flux of protons plus helium at an energy of \SI{100}{TeV}. 
This is also the energy scale where the lowest energy indirect measurements have been made, most recently by HAWC, but previously by ARGO-YBJ. 
If the spectral softening and hardening are universal in \emph{rigidity}, the adding of spectra in \emph{energy} necessarily leads to a certain smoothing of those features. 
Yet, a hardening is still visible around a TeV and a surprisingly sharp softening is seen at a few tens of TeV. 
The measurements by DAMPE show good agreement with HAWC, but given the soft spectrum at the highest energies, another hardening would be needed to connect to the KASCADE data. 

It thus appears that the spectra of CR nuclei at energies between hundreds of \SI{}{\giga\eV} and \SI{\sim 100}{TeV} are much richer than thought. 
This has spawned some interpretations in terms of multiple populations of sources or even particular instances of individual sources.
A number of fits of the spectra in this energy range have thus resorted to fitting it with a combination of broken power laws, exponentially cut-off power laws or log-parabolas, sometimes enforcing relations between the spectral positions of different species implied by universality arguments. 
Mapping such spectral features into a transport model might however be rather challenging. 
If we adopt a simplified picture in which all particles are released from an individual source on a time scale much shorter than the time scales of transport, one can say that the observed spectral shape from an individual source depends on the spectrum of particles released, the source age and the source distance. 
Adopting power law source spectra and diffusion coefficients results in rather broad features which seem not to be able to accommodate some of the very narrow features observed. 
In addition, even if for a population of sources the right ages and sources could be found that allow reproducing the data, it is by no means guaranteed that these parameters could be statistically compatible with a model for the distribution of ages and distances. 
For instance, the fit could require the presence of a very nearby and very young source of CRs while population models might attach a tiny likelihood to such a configuration.
This situation is very much reminiscent of the situation in CR electrons and positrons discussed at the end of Sec.~\ref{sec:interpretation_electrons_positrons}.

% ----------------------------------------------------------------------------------------
% ----------------------------------------------------------------------------------------
% ----------------------------------------------------------------------------------------
\section{Cosmic rays as actors}
\label{sec:actors}

We now turn to the alternative view of CRs mentioned in the introduction, considering the active role that CRs play in shaping the environments they reside in. 
Much attention in that respect is devoted to CR sources and the Galaxy at large. 
The former has been a line of investigation ever since it was realised that the magnetic fields need to be amplified over their ISM values in order to more efficiently confine high-energy particles in shock acceleration. 
The non-resonant hybrid, a.k.a. Bell instability is currently the best candidate for such a mechanism. 
The latter environment has received enhanced attention over the last couple of years, as the importance of CR effects has been realised. 

% ----------------------------------------
% B. Schroer #163
% ----------------------------------------
The role of the non-resonant streaming instability was realised to likely extend beyond the immediate vicinity of the shock in CR sources and to be able to affect the gas dynamics around sources. 
In the so-called bubble scenario~\cite{Schroer:2021iic}, the highest energy CRs accelerated in a source can escape and their current can trigger the non-resonant instability. 
While the growth rate is largest on small scales, subsequently turbulence cascades also to \emph{larger} scales, thus confining also the high-energy particles. 
In one discussion session there was disagreement about the question whether the triggering of the non-resonant instability by the escape current and the self-confinement were in fact in conflict. 
Under the condition that the high-energy CRs are confined, however, their pressure can excavate a bubble, thus significantly reducing the gas density in the vicinity of the source. 
This idea has been investigated by PIC simulations which largely seem to confirm the bubble scenario. 

% ----------------------------------------
% C. Bustard #170
% ----------------------------------------
On somewhat larger scales, the interaction of CRs on individual molecular clouds (MCs) has been studied~\cite{Bustard170}. 
Of course, PIC simulations have trouble covering the relevant spatial and temporal scales. 
Instead, MHD simulations have been used with the addition of streaming CRs. 
Interestingly, the generation of turbulence leads to a drop in the CR pressure inside the MC and in the 1D setup adopted the pressure gradient force can put the molecular cloud into motion. 
Particularly interesting in this respect is the importance of damping processes. 
For instance, once ion-neutral damping is taken into account, the drop of the CR pressure is not taking place in the bulk of the MC, but solely at the interface with the ISM. 

% ----------------------------------------
% P. Girichidis #180
% ----------------------------------------
On even larger scales, the effects of CRs have been investigated on the Galaxy as a whole.
Traditionally, in galaxy simulations, CRs have been treated as a fluid, ignoring essentially the distribution of CRs with energy. 
This can only capture the dynamics of that fraction of CRs that dominate the energy density while the transport of CRs of other energies is poorly modelled, for instance since their diffusivities will be different. 
Recently, several groups have attempted a proper spectral treatment of CRs in galaxy simulations. 
Technically, the CR spectrum can be implemented as a piece-wise power law and diffusion is implemented in such a way that the CR energy density is conserved. 
First simulation results have now become available, showcasing the importance of the proper spectral treatment~\cite{Girichidis180}. 
For instance, the driving of outflows by CRs is much more prominent in the improved simulations compared to the ones with fluid CRs.
Due to the stronger outflows, star formation will be significantly suppressed with potentially far-reaching consequences for galaxy evolution. 

% ----------------------------------------
% T. Thomas #145
% ----------------------------------------
Additionally, the backreaction of CRs on the turbulence needs to be considered~\cite{Thomas145}. 
The challenge for MHD simulations is, of course, to resolve the spatial scales relevant for resonant scattering of GeV particles. 
Certain prescriptions have been suggested for estimating the diffusion coefficient from the energy available for gyro-resonant interactions. 
However, the diffusivities obtained in this approach can be very large indeed. 
Some questions thus likely remain to be answered before such approaches can be routinely used in the phenomenology of Galactic CRs.

% ----------------------------------------------------------------------------------------
% ----------------------------------------------------------------------------------------
% ----------------------------------------------------------------------------------------
\section{The very local ISM}
\label{sec:very_local}

Of course, local measurements of fluxes are not the only observational handle on Galactic CRs. 
Non-thermal electromagnetic radiation, e.g.\ in the radio or $\gamma$-ray bands gets produced from interactions of CRs with matter and radiation fields elsewhere. 
This allows probing the CR spectra elsewhere in the Galaxy. 
This constitutes a rich and exciting research direction in itself and contributions on this are covered by the $\gamma$-ray indirect rapporteur~\cite{Mitchell:2021mrb}. 
Here, we will highlight two contributions that use $\gamma$-ray observations of nearby molecular clouds to investigate the CR fluxes in the very local interstellar medium. 

\begin{figure}
\includegraphics[width=0.54\textwidth]{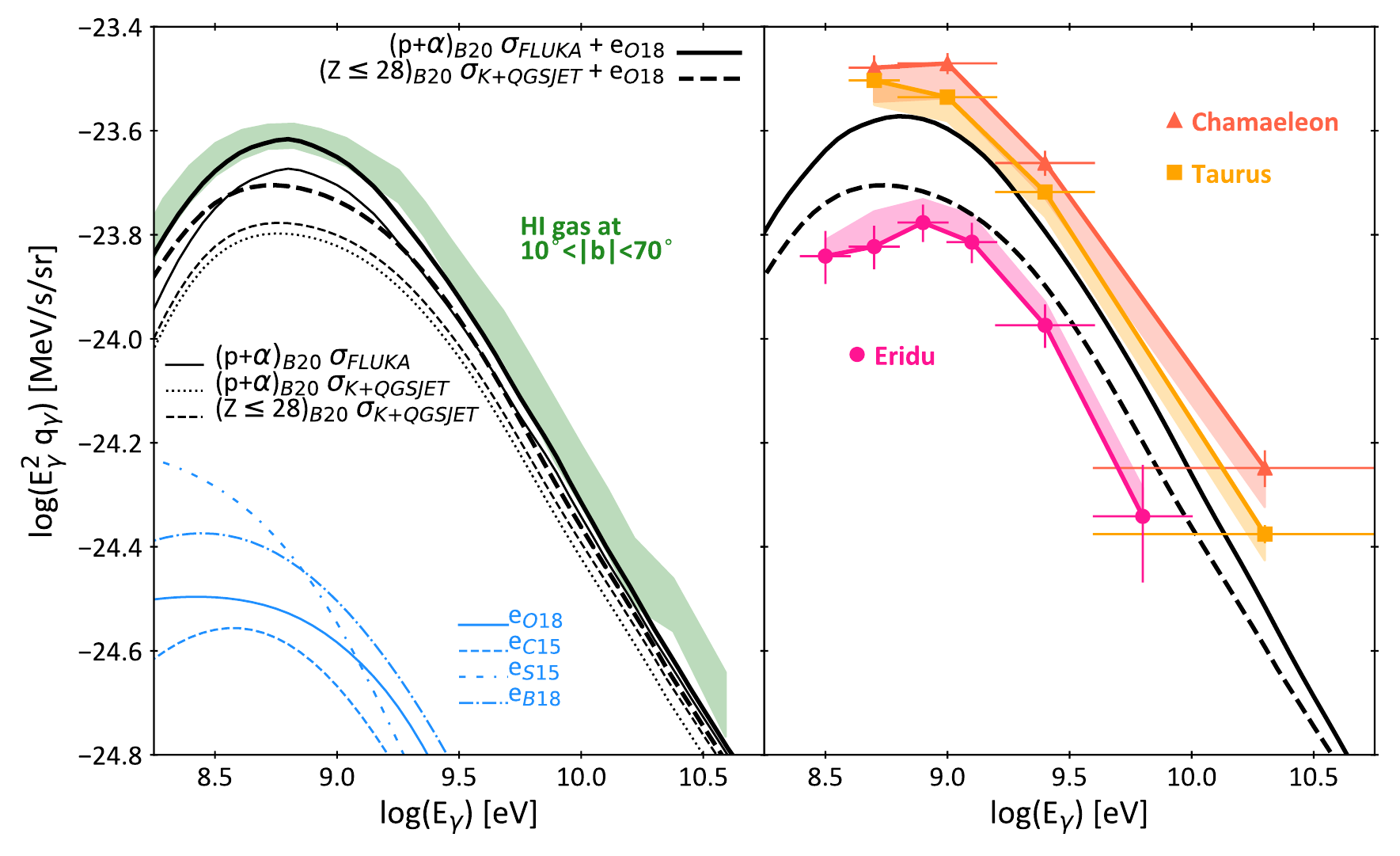}
\includegraphics[width=0.44\textwidth]{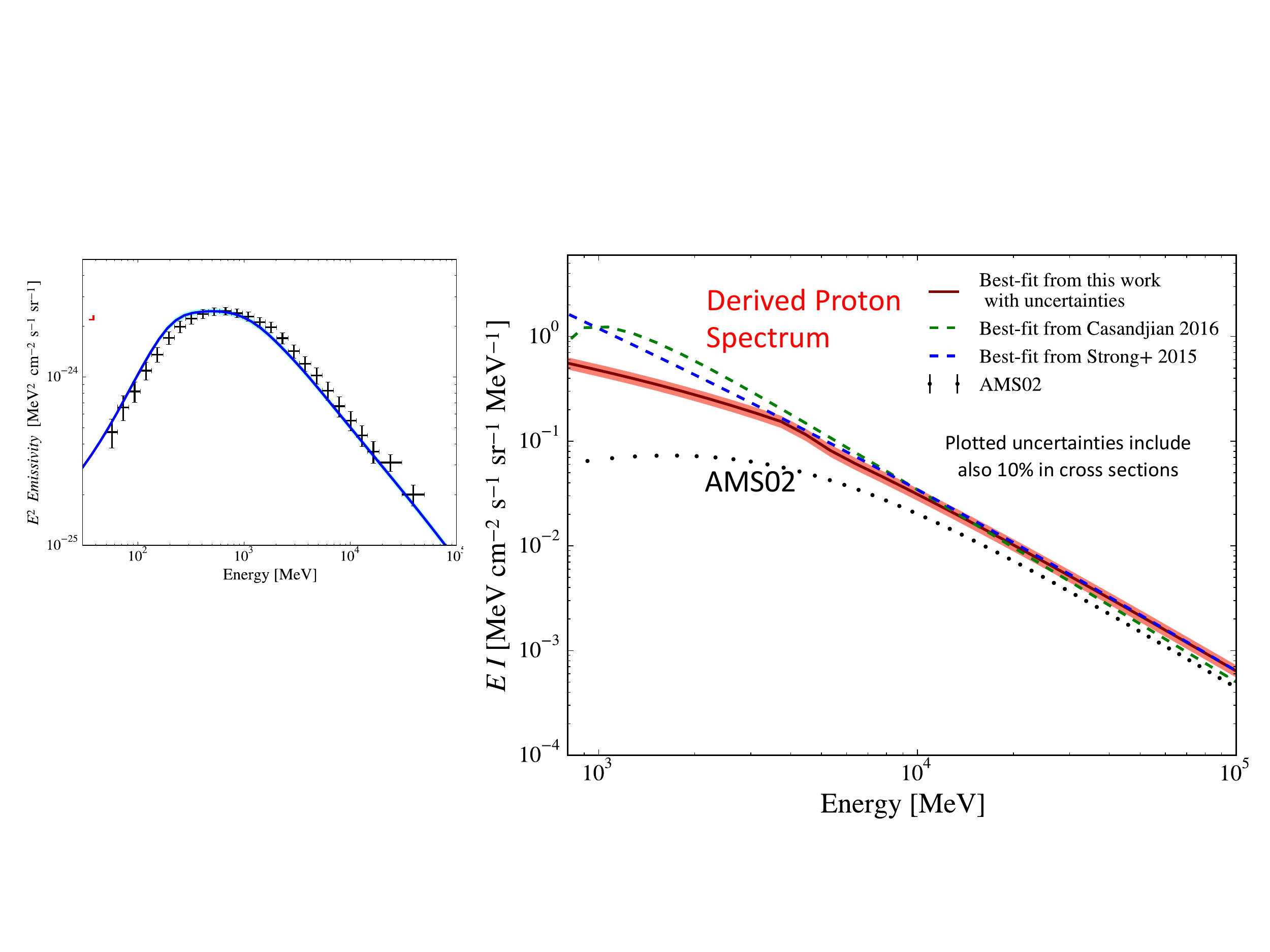}
\caption{Comparison of CR spectra inferred from $\gamma$-rays and local measurements. \textbf{Left panels:} Predictions from different combinations of hadronic models and nuclear models plus inverse-Compton contributions are shown by the different lines, the ranges allowed by data are marked by the coloured bands. From Ref.~\cite{Grenier:2021ban}. \textbf{Right panel:} The proton spectrum inferred from diffuse $\gamma$-ray measurements is shown by the red band, previous analyses by the various dashed lines. The spectrum measured locally by AMS-02 is shown by the dotted line. From Ref.~\cite{Orlando141}.}
\label{fig:diffuse}
\end{figure}

% ----------------------------------------
% I. Grenier #616
% ----------------------------------------
An important input for such studies are of course the $\gamma$-ray production cross-section which are known with less precision that one would hope for. 
In fact, comparing the predictions for the $\gamma$-ray emissivity for a fixed proton flux plus contributions from heavy nuclei and the inverse-Compton scattering of electrons, the predictions differ by \SI{20-30}{\percent} between different hadronic models~\cite{Grenier:2021ban}. 
Yet, comparing the predicted $\gamma$-ray emissivities with measurements, none of the models can explain all the data. 
In the left panel of Fig.~\ref{fig:diffuse}, the $\gamma$-ray emissivity predicted by various hadronic models plus the inverse-Compton contribution are shown. 
This is compared to the emissivity as inferred from high-latitude neutral hydrogen (green band) and also in the left panel to data from two nearby molecular clouds (yellow and orange bands) as well as a nearby cloud with significant fraction of atomic gas (magenta band).

% ----------------------------------------
% E. Orlando #141
% ----------------------------------------
The comparison can of course also be done on the level of the CR fluxes~\cite{Orlando141}. 
The proton spectrum needed to reproduce the emissivity observed again from high latitudes is found to be about \SI{30-40}{\percent} higher than what is found locally by AMS-02.

\begin{figure}
\includegraphics[width=\textwidth,trim={0 0 0 1.5cm},clip=True]{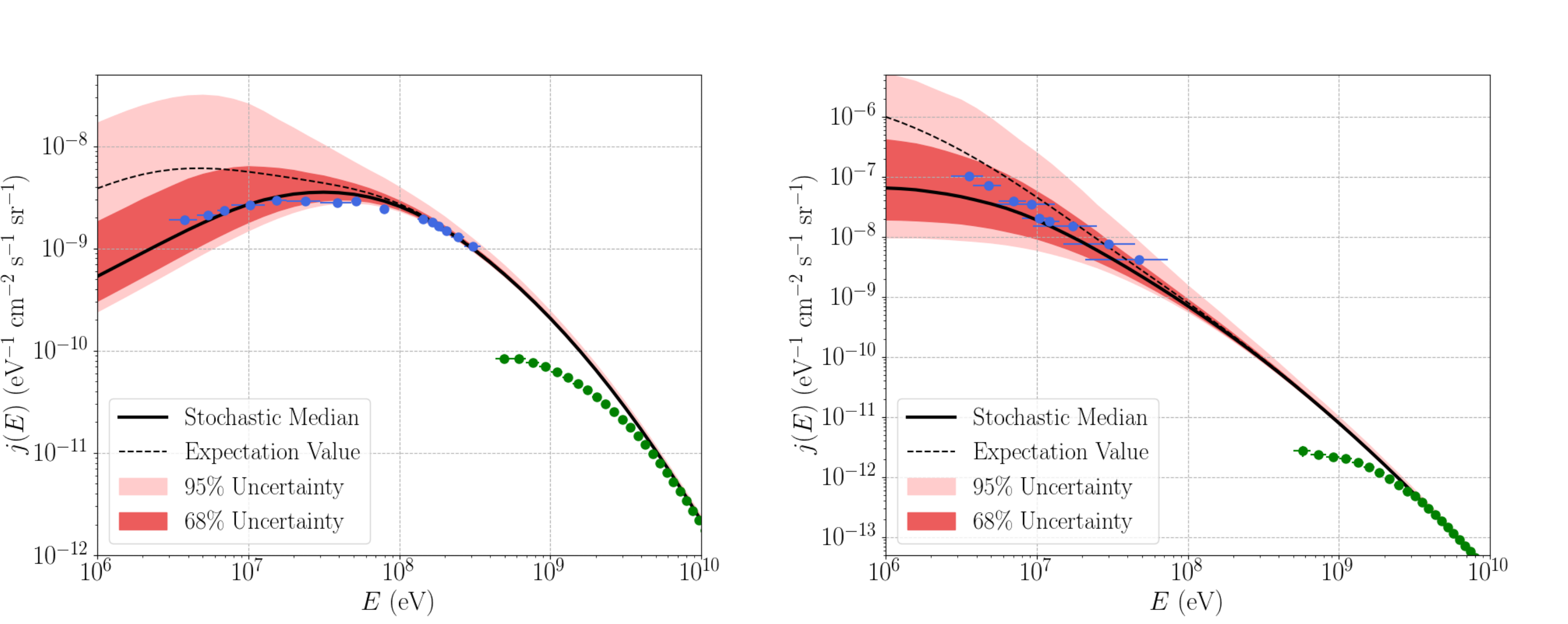}
\caption{Stochasticity in low energy proton (left panel) and electron spectra (right panel). The shaded bands show the 1 and \SI{2}{\sigma} uncertainty bands in an ensemble of sources, the dotted and dashed lines the mean and median of the distribution of fluxes. From Ref.~\cite{Phan165}.}
\label{fig:stochasticity}
\end{figure}

% ----------------------------------------
% M. Phan #165
% ----------------------------------------
This discrepancy, of course, raises the question of how representative the locally measured spectra are, not only for the Galaxy as a whole, but also for our local environment. 
This questions has been studied quantitatively by simulating the CR fluxes contributed by an ensemble of sources~\cite{Phan165}. 
Below \SI{\sim 1}{GeV}, the range of CRs is significantly limited due to ionisation losses and only a small number of sources can contribute. 
Depending on the exact configuration of these sources in space and time, which we have little information on, the fluxes can differ significantly. 
This is illustrated by the shaded bands shown in Fig.~\ref{fig:stochasticity} that are to be interpreted as uncertainty bands of the prediction. 
Interestingly, the statistical mean (dotted line) is not a good estimator of the typical flux from the ensemble and instead the median of the distribution of fluxes should be used. 

% ----------------------------------------------------------------------------------------
% ----------------------------------------------------------------------------------------
% ----------------------------------------------------------------------------------------
\section{Anisotropies}
\label{sec:anisotropies}

Closely related to the presence of nearby sources is the question of CR anisotropies. 
The large-scale anisotropies (``harmonic modulations'') were already studied in the early observations of CRs in extensive air showers. 
Recently there has been renewed interest as anisotropies can contain valuable information on the spatial distribution of CR sources. 
Given the small amplitudes of the dipole anisotropy at energies in reach of direct observations, traditionally, anisotropies have been studied mostly with high-statistics observatories that use indirect measurements. 
More recently, however, direct observatories like AMS-02 have started studying the dipole amplitude and directions at GeV energies.
See Ref.~\cite{Ahlers:2016rox} for a recent review including some observational details in the framework of small-scale anisotropies. 

At this ICRC, the AMS-02 collaboration presented updated constraints on the dipole amplitude in protons and helium as well as for electrons and positrons. 
Given the low levels of anisotropies to be expected, typically of the order $10^{-4 \mathellipsis -3}$ for nuclei, the exposure needs to be known to very high precision. 
The limits are usually presented for a sample integrated in rigidity above a certain minimum value. 
% ----------------------------------------
% M. A. Velasco #108
% ----------------------------------------
No excess has been detected, however, and instead the upper limits are in agreement with the expectations from an isotropic distribution of arrival directions~\cite{VelascoFrutos:2021zjw}.
% ----------------------------------------
% M. Molero #120
% ----------------------------------------
A similar conclusion can be reached for CR electrons and positrons~\cite{MoleroGonzalez:2021con}. 
These species have received additional attention as a probe of the likely nearby source of primary electrons and positrons responsible for the positron excess. 
Among the candidate sources, pulsars have been rather popular and simple isotropic diffusion models have given rather optimistic predictions for the expected size of the dipole amplitude. 
Given the statistics accumulated by AMS-02 so far, the expected limit in the absence of any anisotropy is however still compatible with these optimistic predictions, as is the actual limit. 
Additional statistics during the remaining lifetime of AMS-02 and the ISS will however lead to lower bounds or a detection. 

On the theoretical side, the contributions to this ICRC on CR anisotropies have been using a combination of analytical work and simulations in synthetic turbulence.
% ----------------------------------------
% G. Giacinti #455
% ----------------------------------------
While usually considered a solved problem, it has been pointed out that the large-scale anisotropy need not be a dipole and can instead posses much broader minima and maxima. 
This was based on considering the pitch-angle diffusion coefficient predicted with, e.g.\ a broadened resonance function and has now been confirmed qualitatively by numerical test-particle simulations~\cite{Giacinti:2021jwi}. 

% ----------------------------------------
% Y. Genolini #160
% ----------------------------------------
A conceptually difficult questions is the relation between local and global diffusion. 
It is well know that in the presence of a coherent background field, diffusion is locally anisotropic, meaning that the diagonal components of the diffusion tensor in a field-aligned coordinate system differ. 
However, the particle transport through turbulent magnetic fields is diffusive only in the ensemble average. 
Therefore, when computing diffusion coefficients, for instance through test-particle simulations, averages over many different particle trajectories are computed and hence the effect of the turbulent magnetic field  is averaged over extended spatial regions. 
Computing the local diffusion tensor requires additional effort, but a method for doing this has been presented~\cite{Genolini:2021zio}. 
The results show that generally in the late-time limit all three non-zero components of the diffusion tensor are different. 

% ----------------------------------------
% M. Kuhlen #164
% ----------------------------------------
A longstanding problem is the origin of the small-scale anisotropies observed at TeV and PeV energies by observatories such as IceCube and HAWC. 
While they have been observed in test particle simulations, too, a first principles, analytical treatment was still elusive. 
However, considering the correlated diffusion of pairs of CRs through a turbulent magnetic field, it has been shown that the angular power spectrum from simulations can be reproduced~\cite{Kuhlen:2021qln}. 
The details of this process depend sensitively on the properties of turbulence, as encoded, for instance in the rate of pitch-angle scattering. 
In the future, small-scale anisotropies therefore hold the promise of providing additional handles on the nature of turbulence and its effects on charged particle transport.

% ----------------------------------------------------------------------------------------
% ----------------------------------------------------------------------------------------
% ----------------------------------------------------------------------------------------
\section{Future experiments}
\label{sec:future}

There cannot be any doubt that great progress has been made on the experimental side over the last couple of years. 
Future efforts aim at pushing the maximum attainable energies for direct measurements even further, studying the ultra-heavy composition and achieving unprecedented mass composition. 
% ----------------------------------------
% S.-N. Zhang
% ----------------------------------------
In the near term, the highest energies will be likely achieved by the HERD calorimeter experiment~\cite{Zhang_session_15}, to be installed on the Chinese space station around $\sim$ 2027. 
It promises direct measurements of nuclei between \SI{30}{GeV} and \SI{3}{PeV}, thus providing the first direct measurement at the CR knee. 
Electrons will be measured between \SI{10}{GeV} and \SI{100}{TeV} thus covering the range crucial for the spectral signatures of individual sources. 
% ----------------------------------------
% N. Park # 91
% ----------------------------------------
The balloon-born spectrometer HELIX~\cite{Park:2021oic} aims at measuring the isotopic composition at GV rigidities, e.g.\ the Beryllium ratio. 
It consists of a drift chamber tracker, a time-of-flight system and a ring imaging Cherenkov counter and, given it large size and field of view, it will be able to achieve unprecedentedly small errors on $\mathstrut^9\text{Be} / \mathstrut^{10}\text{Be}$ in a matter of days. 
It is planned to fly in 2022 already. 
% ----------------------------------------
% J. Mitchell #86
% ----------------------------------------
Finally, the TIGERISS experiment~\cite{Rauch:2021akc} is an extension of the successful TIGER and SuperTIGER experiments and will be aimed at measuring all nuclei including the ultra-heavies up to Z=85. 
While still in the planning stages, it would deploy to the Japanese experimental module (JEM) on the ISS. 

For the long term future, even more ambitious projects have been proposed. 
% ----------------------------------------
% S. Schael
% ----------------------------------------
AMS-100~\cite{Schael_session_15} is a large spectrometer to be operated at Lagrange point 2. 
It consists of a solenoidal \SI{1}{T} magnet of \SI{6}{m} length and \SI{2}{\m} radius, a cylindrical tracker and a central, cylindrical calorimeter. 
These parameters allow for a maximum detectable rigidity of \SI{100}{TeV}. 
Its prime targets are electrons, positrons and nuclei, but also antinuclei. 
% ----------------------------------------
% R. Battiston
% ----------------------------------------
A competing proposal for a spectrometer to be operated at Lagrange point 2 is ALDAInO~\cite{Battiston_session_15}. 
The science targets largely agree with those of AMS-100. 
The main components of ALDAInO are also a tracker and a spectrometer, but its MDR of \SI{20}{TV} is somewhat lower.

% ----------------------------------------------------------------------------------------
% ----------------------------------------------------------------------------------------
% ----------------------------------------------------------------------------------------
\section{Summary and conclusion}
\label{sec:summary}

In this report we have reviewed the contributions to the CRD track of the 2021 edition of the ICRC. 
Broadly, we have distinguished between contributions aiming at investigating the spectra of CRs, both through observations and modelling, and studies considering the backreaction of CRs on their respective environments. 
A number of new measurements have been presented, both pushing the boundary of direct measurements in energy and extending our reach in species. 
The picture emerging is one significantly more complicated that thought even a decade ago, with many spectral features between a GeV and the CR knee at a few PeV. 
As the various interpretations of recent CR data have shown, there is no doubt that these new data will be extremely valuable for shedding light on the origin of CRs. 
However, additional input will also be needed in terms of cross-section measurements. 
The measurements of CR spectra in the immediate vicinity of the solar system have received renewed attention and we have started examining how representative they are in a quantitative way. 
Theoretical studies have highlighted the most promising approaches to bridging the gap between first principles PIC simulations and other, mostly MHD approaches studying the dynamics of sources on larger, sometimes dynamical timescales. 
Our understanding of CR anisotropies is ever increasing and we can hope to use this in the future to investigate the properties of turbulence in the ISM. 
Regarding the backreaction of CRs, this has been shown to be important beyond the well-known field amplification in non-relativistic shocks: 
Bubbles can be blown in the source vicinity, molecular clouds can be pushed and the overall evolution of galaxies can be altered. 
It remains to be seen whether all of these promises can already be realised before the next ICRC in 2023, to be held in Osaka, Japan. 
However, if recent history has taught us anything, then it is that the high precision observations will provide an ever richer picture of Galactic CRs.

% ----------------------------------------------------------------------------------------
% ----------------------------------------------------------------------------------------
% ----------------------------------------------------------------------------------------
\bibliographystyle{JHEP_5etal}
\bibliography{ICRC2021_044_Mertsch}

\end{document}